\documentclass[referee]{aa}

\usepackage{graphicx}
\usepackage{natbib}
\usepackage{latexsym}
\usepackage{amsmath}
\usepackage{amssymb}
\usepackage{amstext}
\usepackage{color}
\usepackage{psfrag}
                                                                                
\def\k{km s$^{-1}$}
\def\ks{km s$^{-1}$~}
\def\d{$^\circ$}
\def\m{$^\prime$}
\def\s{$^{\prime\prime}$}

\def\cm3{cm$^{-3}$}

\def\2{$^{12}$CO}
\def\3{$^{13}$CO}
\def\H{HCO$^{+}$}
\def\msol{M$_\odot$}

\def\ir{IRAS~18544+0112}

\begin{document}

\title{A multiwavelength study of the star forming region  IRAS 18544+0112}

\author {M. E. Ortega \inst{1}
\and S. Paron \inst{1}
\and S. Cichowolski  \inst{1}
\and M. Rubio  \inst{2}
\and G. Castelletti \inst{1}
\and G. Dubner \inst{1}
}

\institute{Instituto de Astronom\'{\i}a y F\'{\i}sica del Espacio (IAFE),
             CC 67, Suc. 28, 1428 Buenos Aires, Argentina\\
             \email{mortega@iafe.uba.ar}
\and Departamento de Astronom\'{\i}a, Universidad de Chile, Casilla 36-D,
        Santiago, Chile}

\offprints{M.E. Ortega, \email{mortega@iafe.uba.ar}}

\date{Received / Accepted }

\abstract{} {This work aims at investigating the molecular and
infrared components in the massive young stellar object (MYSO)
candidate IRAS~18544+0112. The purpose is to determine the nature and
the origin of this infrared source.}  {To analyze the molecular gas
towards IRAS~18544+0112, we have carried out observations in a 90\s
$\times$ 90\s~ region around $l =$ 34$\fdg$69, $b =$ --0$\fdg$65,
using the Atacama Submillimeter Telescope Experiment (ASTE) in the \2
J=3--2, \3 J=3--2, \H J=4--3 and CS J=7--6 lines with an angular
resolution of 22\s. The infrared emission in the area has been
analyzed using 2MASS and Spitzer public data.} {From the molecular
analysis, we find self-absorbed \2 J=3--2 profiles, which are typical
in star forming regions, but we do not find any evidence of
outflow activity. Moreover, we do not detect either \H J=4--3 or CS
J=7--6 in the region, which are species normally enhanced in molecular outflows
and high density envelopes.  The \2 J=3--2 emission profile suggests
the presence of expanding gas in the region.  The Spitzer images
reveal that the infrared source has a conspicuous extended emission
bright at 8 $\mu$m with an evident shell-like morphology of $\sim$
1\farcm5 in size ($\sim$ 1.4 pc at the proposed distance of 3 kpc)
that encircles the 24 $\mu$m emission. The non-detection of ionized
gas related to \ir\, together with the fact that it is still embedded
in a molecular clump suggest that \ir\, has not reached the UCHII
region stage yet.  Based on near infrared photometry we search for
YSO candidates in the region and propos that 2MASS 18565878+0116233
is the infrared point source associated with \ir. Finally, we suggest
that the expansion of a larger nearby HII region, G034.8$-$0.7, might
be related to the formation of \ir.}  {}

\keywords{ISM: molecules - HII regions - Stars: formation}

\titlerunning{A
multiwavelength study towards IRAS 18544+0112}
\authorrunning{M. E. Ortega et al.}  

\maketitle

\section{Introduction}

Star formation processes start when a pressure-bounded,
self-gravitating molecular clump becomes gravitationally unstable. As
summarized by \citet{wit94a, wit94b}, the action of an expanding
nebula can produce gravitationally unstable shocked layers of
interstellar gas.  For example, the expansion of an HII region can
sweep up the surrounding molecular gas into a dense shell, which then
fragments and forms new massive stars (``collect and collapse'' model;
see \citealt{elme77}).  Several recent works support this model
(e.g. \citealt{deha05,comeron05,zav06,poma09}). On the other hand,
shockwaves from expanding wind- and/or supernova- driven superbubbles
can also trigger cloud collapse and star formation, but at larger
scales. Numerical studies \citep{vanha98,melioli06} demonstrated that
the effect caused by a passing shockwave mainly depends on the
shock-type: close to the supernova remnant (SNR) the shockwave
disrupts the ambient molecular clouds and thus terminates the star
formation process; however, a little further away from the SNR the
shock velocity decreases, and cloud collapse is possible if the right
circumstances were given.

This work is part of a systematic study towards IR sources embedded in
molecular condensations with evidence of being affected by SNRs shock
fronts or expanding HII regions. In a previous work, \citet{paron09}
studied the infrared (IR) source IRAS~18542+0114 located near the
border of the SNR W44. They discovered that this source is probably a
massive young stellar object (MYSO) located at the border of the HII
region G034.8$-$0.7 which is evolving within a molecular cloud.  In
this work, we present a study of the neighboring source
IRAS~18544+0112, another IR source embedded in the same molecular
cloud.  It is important to remark that the molecular cloud, the HII
region G034.8$-$0.7 and the SNR W44 are not only located in the same
region in the plane of the sky but also at the same distance from the
Sun, at about 3 kpc (corresponding to the kinematic velocity of
$v_{LSR} \sim 45$ \k, \citealt{paron09} and references therein).

The source investigated in this Paper, IRAS~18544+0112, was cataloged
as a MYSO and a high mass protostellar object (HMPO) by
\citet{molinari96} and \citet{kum07}, respectively. On the basis of
the analysis of new molecular data and near- and mid-IR data, we present a multiwavelength study of the IR 
source aiming to discern its origin and evolutionary stage.

\begin{figure}
\centering
\includegraphics[width=8cm]{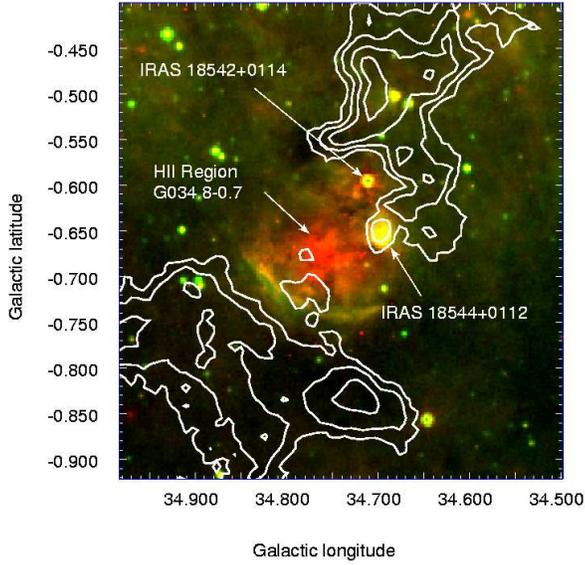}
\caption{Two-color image of the HII region G034.8$-$0.7 containing
IRAS~18542+0114 and IRAS~18544+0112. Green is the {\it Spitzer}-IRAC 8
$\mu$m emission and red is the {\it Spitzer}-MIPSGAL 24 $\mu$m
emission. White contours at 0.4, 0.6, 1 and 1.2 K represent the \3
J=1--0 line emission distribution integrated between the velocity
range from 36 to 45 \k.}
\label{IR824CO}
\end{figure}

\section{IRAS~18544+0112 and its environment}

Figure \ref{IR824CO} shows a two-color IR image of the HII region
G034.8$-$0.7. We show in green the {\it Spitzer}-IRAC 8 $\mu$m
emission and in red the {\it Spitzer}-MIPSGAL 24 $\mu$m emission.
Yellow corresponds to regions where both emissions overlap. The white
contours represent the \3 J=1--0 line emission distribution (as
extracted from the Galactic Ring Survey (GRS;
\citealt{jackson06}) averaged over the velocity interval from 36 to 45
\k.
  
The 8 $\mu$m emission, which arises mainly from the polycyclic aromatic
hydrocarbons molecules (PAHs; \citealt{leger84}) and an underlying
continuum attributed to very small grains, is observed enclosing the 24 $\mu$m
emission that is originated in the heated dust of the HII region
G034.8$-$0.7. These molecules  cannot survive inside HII regions and are located
over the photodissociation regions (PDRs) that encircle the ionized
gas \citep{cesarsky96}. The PDRs are the interphase zone between the
ionized and molecular gas, and its presence evidences the interaction
between them. A well defined 8 $\mu$m arc-like structure is observed
bordering the HII region G034.8$-$0.7 towards lower Galactic
latitudes.  It is important to note that the ionized region is
partially bordered by two molecular clouds which are part of the giant
molecular complex (GMC) G34.8$-$0.6. The morphology of the molecular
clouds, as seen in the \3 J=1--0 line, suggests that they are 
shaped by the action of the HII region.

On the other hand, IRAS~18544+0112, located at ($l, b$) = (34$\fdg$69,
--0$\fdg$65), appears as a bright yellow knot. From Fig.~\ref{IR824CO}
 it can be noticed that this IR source is located in a bulge shaped
molecular gas condensation which is observed in the velocity interval from 36
to 45 \k. 

\begin{figure}
\centering
\includegraphics[width=8cm]{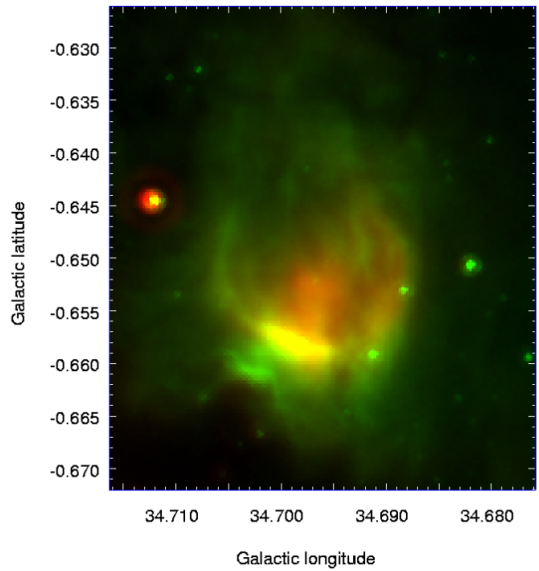}
\caption{The green is the {\it Spitzer}-IRAC 8 $\mu$m emission and the
red is the MIPSGAL emission at 24 $\mu$m towards IRAS~18544+0112.}
\label{8+24}
\end{figure}

Figure \ref{8+24} shows a two-color image of \ir\,, where green and
red are the 8 $\mu$m and 24 $\mu$m emission, respectively.  The
observed 8 $\mu$m emission distribution shows a shell-like structure
of about 1$\farcm$5 in size ($\sim$ 1.4 pc at a distance of 3 kpc),
bright filaments and diffuse emission.  The emission at 24 $\mu$m is
observed mainly inside the shell-like structure.

Since most HMPOs present molecular outflows, we now analyze whether this
phenomenon is taking place in the region.

\begin{figure}
\centering \includegraphics[width=9cm]{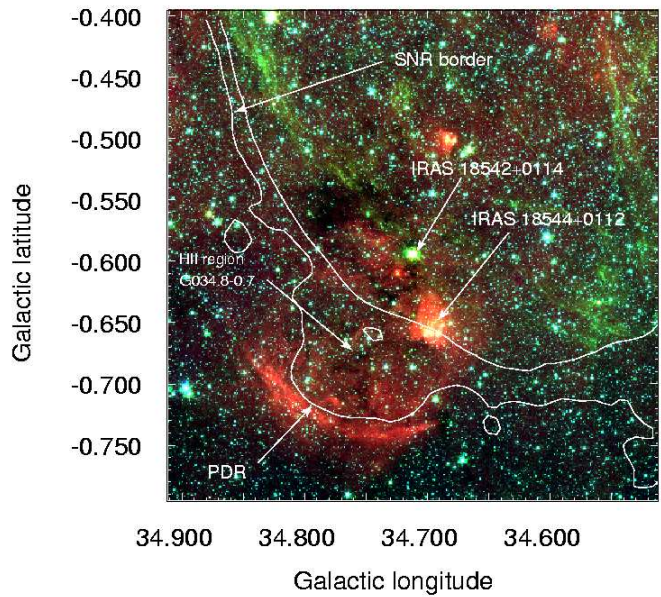}
\caption{{\it Spitzer}-IRAC three-color image (3.5 $\mu$m $=$ blue,
  4.5 $\mu$m $=$ green, and 8 $\mu$m $=$ red) of a region of 22\m\, in
  size. The contours correspond to the 1.4 GHz radio continuum
  emission at 17 and 35 K, which depict the border of the HII region
  G034.8$-$0.7 and the SE border of the SNR W44, respectively.}
\label{GLIMPSE3colors}
\end{figure}

The IRAC 4.5 $\mu$m band contains lines that may be excited by high
velocity shocks, such as those expected when protostellar outflows
crash into the ambient ISM \citep{cyga08} or when a SNR interacts with
a molecular cloud. We thus inspected the IRAC 4.5 $\mu$m emission
distribution in the area.

Figure \ref{GLIMPSE3colors} shows a {\it Spitzer}-IRAC three-color
image of a square region of 22\m~in size. The three IR bands presented
are: 3.6 $\mu$m (in blue), 4.5 $\mu$m (in green) and 8 $\mu$m (in
red). The white contours represent the radio continuum emission at 1.4
GHz as extracted from the VGPS (VLA Galactic Plane Survey;
\citealt{sti06}), which depict the SE border of the
SNR W44 and the border of the radio continuum emission associated with
the HII region G034.8$-$0.7 encompassing the PDR.

An inspection of Fig.~\ref{GLIMPSE3colors} shows that the source
IRAS~18542+0114 studied by \citet{paron09}, appears slightly extended
in the 4.5 $\mu$m emission (green). Such a characteristic is suggestive
of a YSO origin for the emission, a condition that was confirmed by
\citet{paron09} after discovering associated molecular
outflows. Figure ~\ref{GLIMPSE3colors} also shows some extended
diffuse filaments in this IRAC band, which are probably illuminating
molecular gas shocked by the SNR.  In the case of the source studied
in this work, IRAS~18544+0112, it does not present significant
emission in the 4.5 $\mu$m band. Indeed, this source is clearly
brighter at 8 $\mu$m than at 4.5 $\mu$m, suggesting that there is no
outflow activity in the region.

With the purpose to analyze the small scale distribution and
dynamic of the molecular gas associated with \ir\,, we carried out
observations of the \2, \3 J=3-2, \H J=4-3, and CS J=7-6 lines towards
a region of 90\s $\times$90\s~around this IR source using the Atacama
Submillimeter Telescope Experiment (ASTE; \citealt{ezawa04}).

\section{New molecular observations}

The molecular observations were performed on June 25, 2008 with the 10
m ASTE Telescope. We used the CATS345 GHz band receiver, which is a
two-single band SIS receiver remotely tunable in the LO frequency
range of 324-372 GHz. We simultaneously observed \2 J=3--2 at 345.796
GHz and \H~J=4--3 at 356.734 GHz, mapping a region of 90\s~$\times$
90\s~centered at the position of IRAS 18544+0112, ($l, b$) =
(34$\fdg$69, --0$\fdg$65). The mapping grid spacing was 10\s~and the
integration time was 72 sec. per pointing. Additionally, we observed \3
J=3--2 at 330.588 GHz and CS J=7--6 at 342.883 GHz towards the center
of the region. All the observations were performed in position-switching mode. The off-position was ($l, b$) = (34$\fdg$87,
--0$\fdg$14), which was checked to be free of emission.

We used the XF digital spectrometer with a bandwidth and spectral
resolution set to 128 MHz and 125 kHz, respectively.  The velocity
resolution was 0.11 \ks and the half-power beamwidth (HPBW) was
22\s~at 345 GHz. The system temperature varied from T$_{\rm sys} =
400$ to 700 K. The typical rms noise (in units of T$_{\rm mb}$) ranged
between 0.1 and 0.4 K, and the main beam efficiency was $\eta_{\rm mb}
\sim 0.65$.

The spectra were Hanning-smoothed to improve the signal-to-noise ratio,
and only linear or/and some third order polinomia were used for
baseline fitting. The spectra were processed using the XSpec software
package developed at the Onsala Space Observatory.  

\section {Molecular analysis}

As it was shown in Fig. \ref{IR824CO}, IRAS~18544+0112 appears
embedded in a molecular clump, which is part of the GMC G34.8$-$0.6.

\begin{figure}
\centering \includegraphics[width=11cm]{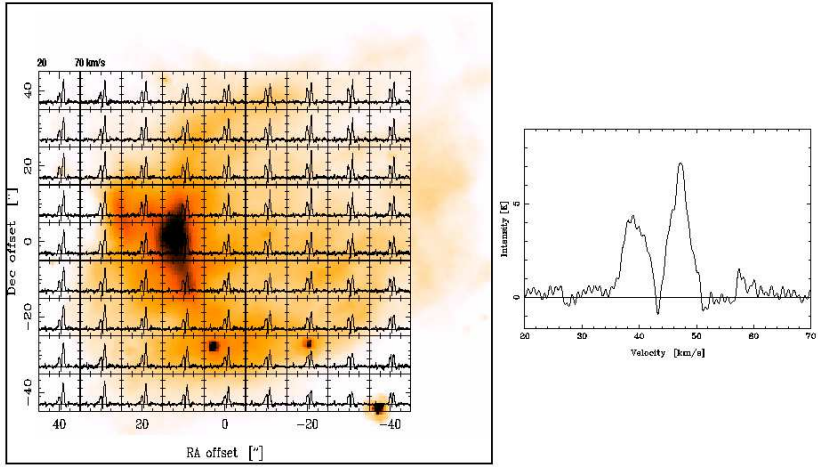}
\caption{{\it Left}: \2 J=3--2 spectra obtained towards IRAS
18544+0112. The velocity range of each spectrum is between 20 and 70
\k.  The color scale corresponds to the emission at 8$\mu$m
associated with \ir\,, as extracted from {\it Spitzer}-IRAC. {\it
Right}: \2 J=3--2 spectrum towards the center of the region.}
\label{CO32spectra}
\end{figure}

\begin{figure}
\centering \includegraphics[width=5cm,angle=270]{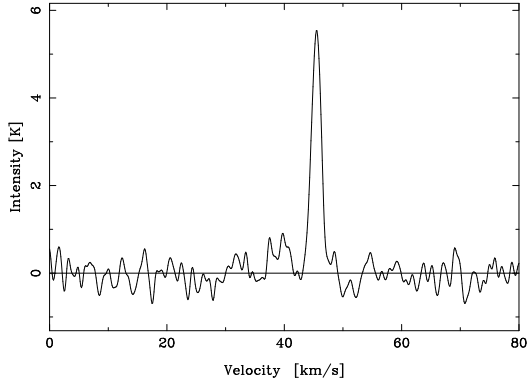}
\caption{\3 J=3--2 profile obtained towards the center of IRAS 18544+0112.}
\label{13CO32}
\end{figure}

Figure~\ref{CO32spectra}\,(left) shows the \2 J=3--2 spectra obtained
from a region of 90\s~$\times$ 90\s~centered at the position of \ir\,
and Fig.\ref{CO32spectra}\,(right) presents a spectrum obtained
towards the center of the region. Figure ~\ref{13CO32} displays an
average spectrum of \3 J=3--2 obtained towards the center of the clump
at ($l, b$) = (34$\fdg$69, --0$\fdg$65). All the \2 profiles in this
region present a dip at $v \sim 44$ \k.  Such a narrow dip points to a
self-absorption origin instead of two \2 emission components with very
close kinematical velocities.  We notice that the $v \sim 44$ \ks dip
is very close to the peak velocity of the \3 J=3--2 line, which is an
optically thinner line. Such correspondence strongly suggests that the
dip in the \2 profiles is in fact caused by self-absorption by less
excited gas (see for example \citealt{zhou93}).

Since we are searching for indicators of active star formation in this
IR source, the presence of this dip is interesting because it is known
that the \2 J=3--2 line has almost always  been observed self-absorbed
in star-forming regions \citep{johnstone03}.

Additionally, we note that the \2 J=3--2 emission does not show any
evidence of outflow activity in the region, which agrees with the
faint 4.5 $\mu$m emission mentioned in Sect. 2. The non-detection of
the \H J=4--3 and CS J=7--6, molecular species that are enhanced in
the interphase layer between the outflows and the surrounding
molecular core where a YSO is forming \citep{hoger98,raw04} points in
the same direction.

On the other hand, Fig.~\ref{CO32spectra} clearly shows that most of
the \2 profiles in the region present a redshifted component brighter
than the blueshifted one, suggesting that the molecular gas is
expanding. This is because in an expanding cloud a line emission is
composed by red and blueshifted photons.  In the case of an optically
thick line, such as the \2 J=3--2, the redshifted photons will
encounter fewer absorbing material (which is expanding outward) than
would blueshifted photons and hence have greater probabilities of
escape (e.g. \citealt{leung78,zhou92,lehtinen97}).

To estimate the physical parameters of the molecular clump where 
IRAS 18544+0112 is embedded, we determine the line parameters of 
the \3 J=3--2 line from a Gaussian fitting. We obtain:
T$_{mb} \sim$ ($5.7 \pm 0.5$)K, V$_{LSR} \sim$ ($45.4 \pm 0.4$)\k,
$\Delta v \sim$ ($2.0 \pm 0.2$) \ks and $I \sim$ ($12 \pm 1$)
K~\k, where T$_{mb}$ represents the peak brightness temperature, V$_{LSR}$
the central velocity referring to the local standard of rest, 
$\Delta v$ the line width and $I$ the integrated
line intensity. Errors are a formal 1$\sigma$ value for the model of the
Gaussian line shape.

Using the \3 J=3--2 and J=1--0 lines we calculated the ratio of the
integrated line intensities ($^{13}$R$_{3-2/1-0}$). The \3 J=1--0 line
was extracted from the GRS and the J=3--2 line was convolved to the
J=1--0 beam. We obtained  $^{13}$R$_{3-2/1-0} \sim 1$.

Assuming LTE conditions and a beam filling factor of 1, which of
course may not be completely true but allows us to make an initial
guess, we use the \2 J=3--2 emission towards the center of the
analyzed region to estimate an excitation temperature, T$_{\rm ex}$.
As noticed above, this emission appears self-absorbed, showing a dip
between the blue and red emission components. Thus, we use an average
between the T$_{mb}$ of both components to obtain T$_{\rm ex}$ $\sim$
15 K. Using this factor and the parameters obtained for the \3 J=3--2,
we derive an optical depth for this line of $\tau^{13}$
$\sim$ 0.2 and a \3 column density of N(\3) $\sim$ 2.2 $\times$
10$^{15}$ cm$^{-2}$. Adopting the $^{12}$CO and $^{13}$CO
relationships of N(H$_2$)/N($^{12}$CO)= 10$^5$ and
N($^{12}$CO)/N($^{13}$CO)= 89 \citep{guan08} and taking into account
that $^{13}$R$_{3-2/1-0}$ = 1 as estimated above, we obtain an H$_2$
column density of N(H$_2$) $\sim$ 2.0 $\times$ 10$^{21}$ cm$^{-2}$.
Finally, assuming a spherical geometry for the clump as seen in the \3 J=1--0
line (Fig.~\ref{IR824CO}-left) with a radius of $\sim$ 0.7 pc, we
estimate a mass and a volume density of $\sim$ 1.7 $\times$ 10$^2$ \msol,
and $\sim$ 3.2 $\times$ 10$^{3}$ cm$^{-3}$, respectively.

\section{Is \ir\, an ultracompact HII region?}

In view of the lack of outflow activity in the area, a possible
explanation is that \ir\, is an evolved high-massive protostellar
object that has finished its accretion phase. Since the
next step in the life of these objects is to develop an
ultracompact HII (UCHII) region we looked for the presence of ionized
gas. The fact that this source lies in the same region of the sky than
the SNR W44 and the HII region G\,34.7--0.8, makes it difficult to
discern whether there is radio emission related to \ir.  An inspection
of the 1420 MHz image obtained from the VGPS does shows the presence of
emission at the position of \ir\,, but its extended morphology suggests
that it is most probably related to G\,34.7--0.8.  To avoid the
contamination of extended radio sources, we have inspected the 1420
MHz radio continuum image obtained from the NRAO VLA Sky Survey (NVSS;
\citealt{con98}). No radio continuum emission is detected at the
position of \ir.  This agrees with the null detection
reported by \citet{hug94} in the direction of this source.

From the cataloged IRAS fluxes of IRAS\,18544+0112 and assuming a
distance of 3 kpc, we estimated the corresponding IR luminosity
L$_{IR}$ and dust temperature T$_d$. The IR luminosity was
estimated on the basis of the four-bands IRAS measurements (12, 25, 60
and 100 $\mu$m) following \citet{chan95} as L$_{IR}$(L$_\odot$)= 1.58
S$_{IR}$ (Jy) D$^2$ (kpc), where S$_{IR}$ is the integrated flux given
by S$_{IR}$ =
1.3(S$_{12}$+S$_{25}$)+0.7(S$_{25}$+S$_{60}$)+0.2(S$_{60}$+S$_{100}$)
and $S_i$ is the flux density in the IRAS band $i$ expressed in Jy. We
obtained L$_{IR}$ $\sim$ (7.9 $\pm$ 1.6) $\times$ 10$^3$ L$_\odot$.

Adopting standard parameters for dust grains \citep{dra84}, the dust
temperature can be derived from the relation T$_d$(K) = (95.94/ln
B$_n$), where B$_n$ = 1.667$^{3+n}$ S$_{100}$/S$_{60}$ is the modified
Planck function, and n is a parameter related to the absorption
efficiency of the dust ($\kappa_{\nu}~\alpha~\nu^n$). We obtained
T$_d$ $\sim$ (29.0 $\pm$ 5.0) K for the adopted value n=1. The
estimated dust temperature of about 29 K for IRAS 18544+0112 is quite
low for a UCHII region, which typically has temperatures of about 200
K \citep{bal96}, while it agrees with the ones derived for
high mass protostellar candidates by \citet{sri02}.

In summary, from the observed characteristics we suggest that \ir\, is
not yet a UCHII region.

\section{A search of MYSO candidates in IRAS 18544+0112}

To search for MYSO candidates associated with \ir\, we performed
a near-infrared photometric analysis of all the sources that are
enclosed within the borders of the IR source as seen in the 8 $\mu$m
band (see e.g. Figure \ref{8+24}).  We used the 2MASS All-Sky Point
Source Catalogue \citep{skr06} in bands {\it J} (1.25 $\mu$m),
{\it H} (1.65 $\mu$m) and {\it K$_{S}$} (2.17 $\mu$m), selecting only
the sources detected in at least two bands. We found 13 sources in
this region which are listed in Table \ref{sources} and shown in
Fig. \ref{sourcesfig} (a) - (c).  In Table \ref{sources} we present
the source numbers (same as in the following figures), the Galactic
coordinates in degrees, the Two Micron All Sky Survey (2MASS)
designation and the {\it J}, {\it H} and {\it K$_{S}$} magnitudes in
cols. 1 to 7, respectively.  Figure \ref{sourcesfig} (a) - (c)
displays the spatial location of these sources superimposed over: the
8 $\mu$m emission (a), the near infrared {\it JHK} three-color image
extracted from the 2MASS (b), and the optical emission as extracted
from the 2nd Digitized Sky Survey Blue (DSS2-B) (c). The dashed circle
represents the area in which we searched for the mentioned sources.

\begin{table}
\caption{2MASS sources towards IRAS 18544+0112.}
\label{sources}
\centering
\begin{tabular}{ccccccc}
\hline\hline
No. & $l$        & $b$        & 2MASS             & $J$     & $H$     & $K_{S}$    \\
\hline 
1   & 34.69  & -0.64  & 18565828+0116334  & 15.73 & 15.18 & 14.37      \\   
2   & 34.70  & -0.65  & 18570030+0116478  & 18.59 & 15.65 & 13.35       \\  
3   & 34.69  & -0.65  & 18565878+0116233  & 17.76 & 14.59 & 12.37       \\  
4   & 34.69  & -0.65  & 18565848+0116162  & 17.04 & 15.89 & 15.13       \\  
5   & 34.69  & -0.65  & 18565987+0116116  & 16.71 & 15.71 & 15.66       \\          
6   & 34.68  & -0.65  & 18565808+0115548  & 15.27 & 11.58 &  9.76       \\
7   & 34.69  & -0.65  & 18565816+0116025  & 18.04 & 15.69 & 13.97       \\          
8   & 34.69  & -0.64  & 18565802+0116298  & 14.44 & 13.86 & 13.70      \\
9   & 34.69  & -0.64  & 18565665+0116434  & 17.46 & 13.94 & 12.10      \\          
10  & 34.70  & -0.65  & 18565913+0116360  & 14.25 & 13.78 & 13.55      \\          
11  & 34.70  & -0.64  & 18565855+0116541  & 17.14 & 15.31 & 14.50       \\       
12  & 34.69  & -0.64  & 18565794+0116215  & 16.10 & 15.19 & 14.21       \\      
13  & 34.69  & -0.65  & 18565971+0115541  & 15.11 & 11.38 &  9.49       \\
\hline
\end{tabular}
\end{table}

\begin{figure}
\centering
\includegraphics[width=6.5cm]{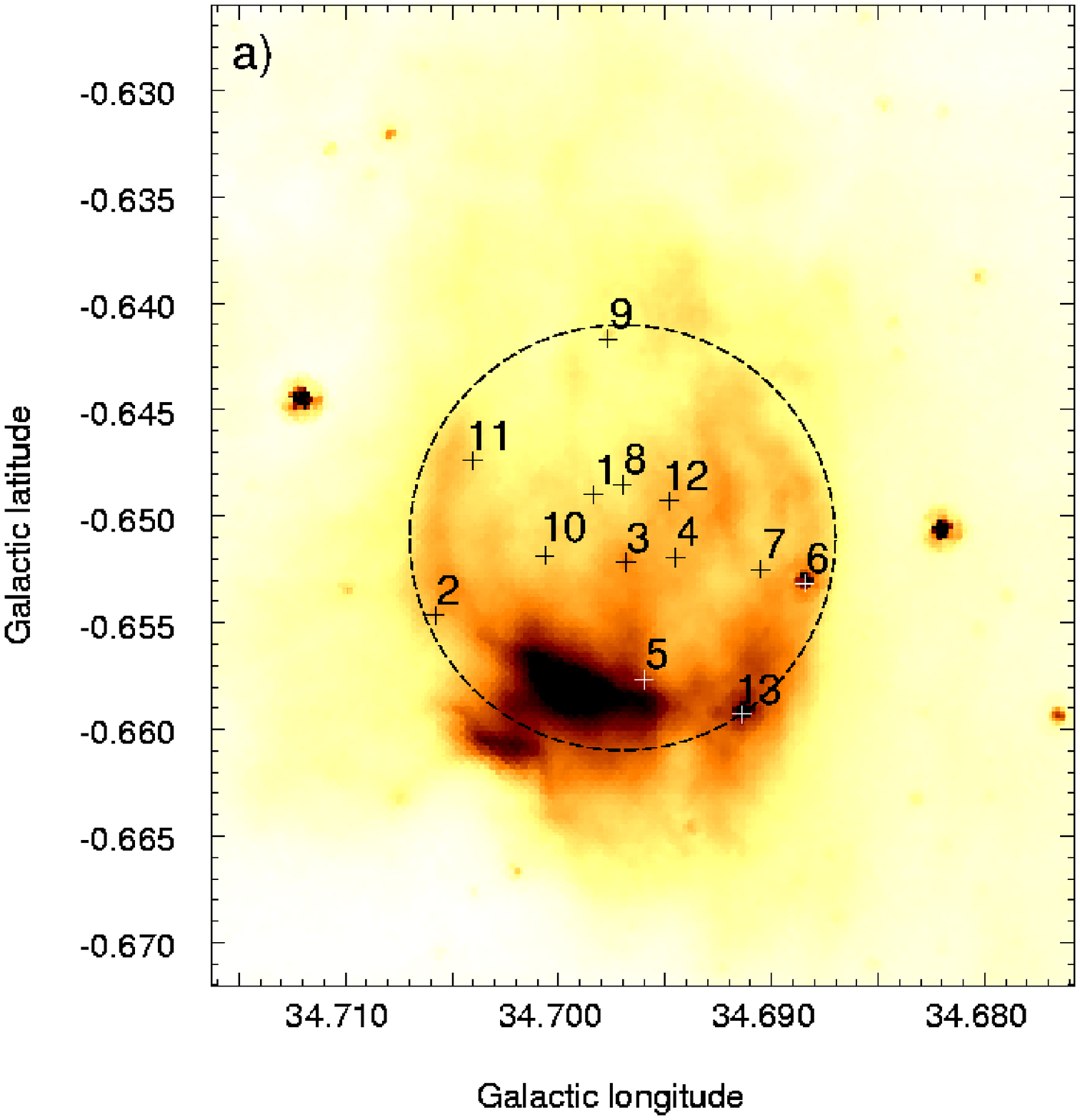}
\includegraphics[width=6.5cm]{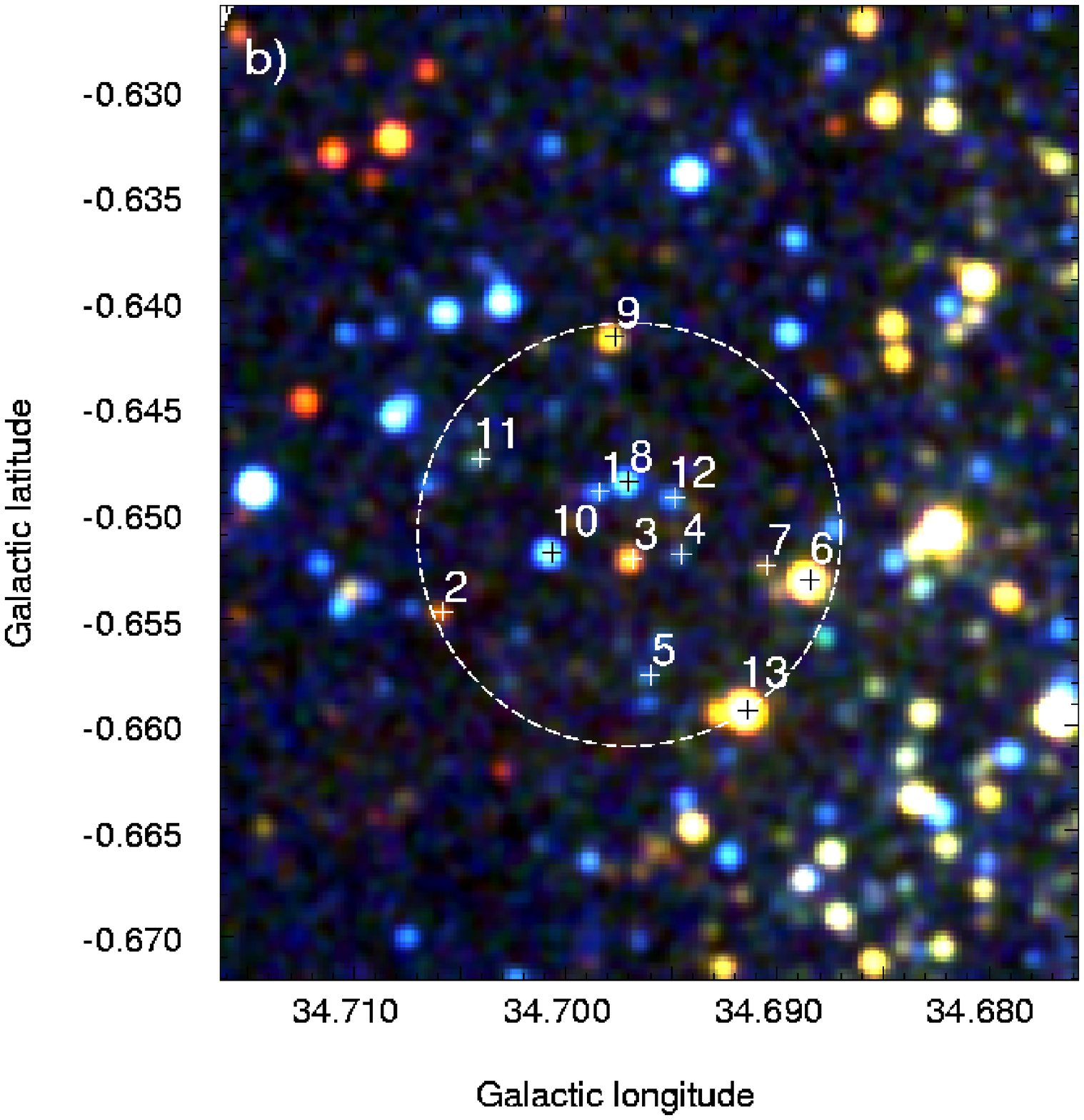}
\includegraphics[width=6.5cm]{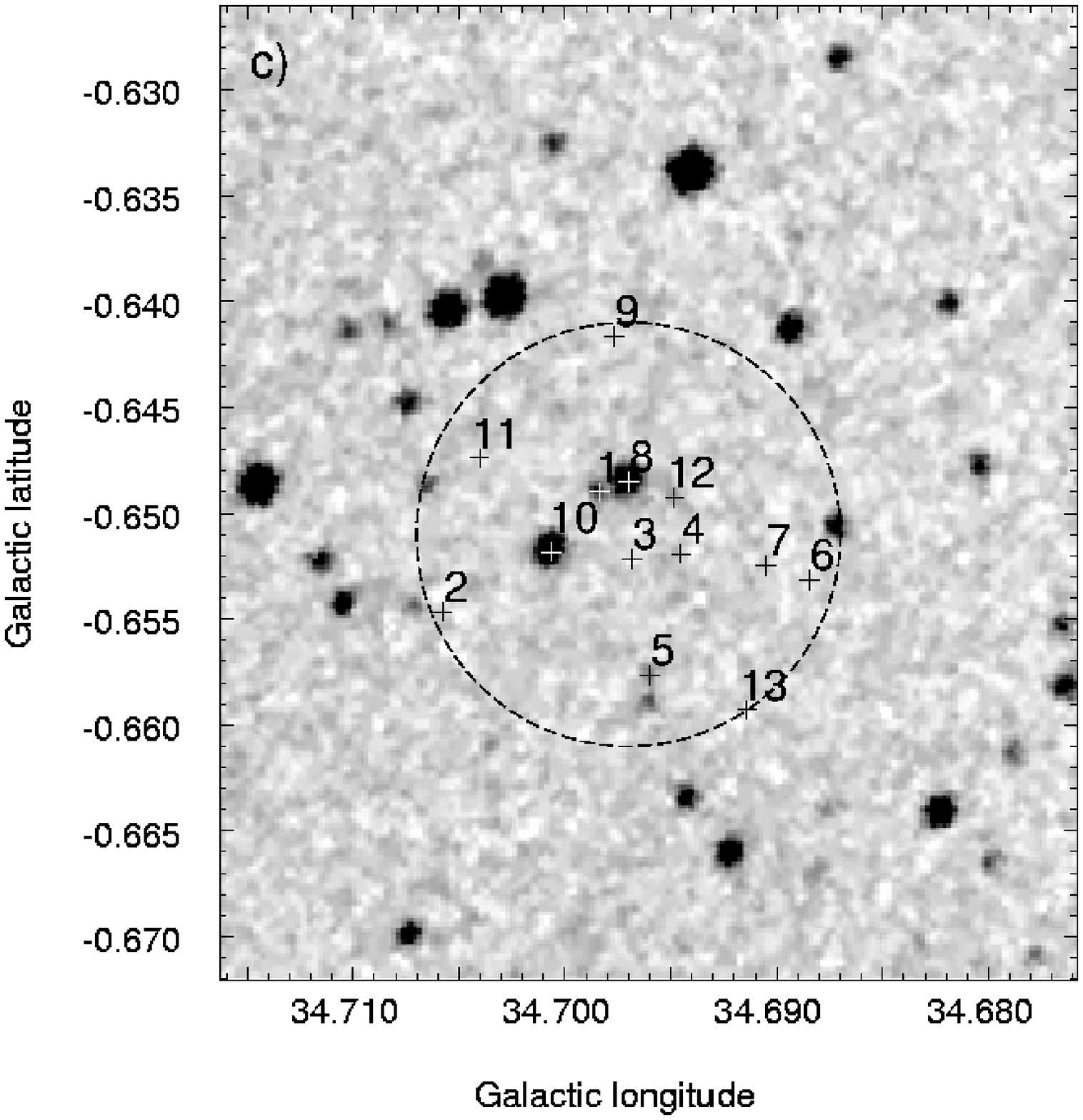}
\caption{Spatial location of the 2MASS sources found towards IRAS
18544+0112 over the 8 $\mu$m emission (a), the near infrared {\it JHK}
three-color image extracted from The Two Micron All Sky Survey (b) and
the DSS2-B optical image (c). The dashed circle represents the area in
which we searched for the mentioned sources.}
\label{sourcesfig}
\end{figure}

\begin{figure}
\centering
\includegraphics[totalheight=0.25\textheight,viewport=0 0 700
450,clip]{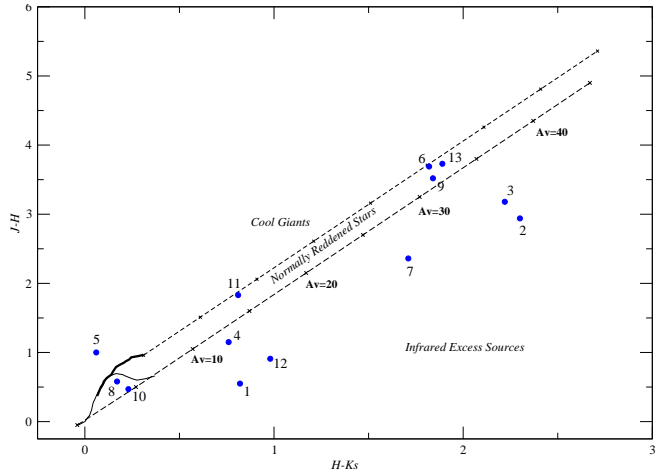}
\caption{Color-color diagram of the 13 2MASS sources found towards IRAS 18544+0112. 
The two solid curves represent the location of the
main sequence (thin line) and the giant stars (thicker line) derived
from \citet{bessell88}. The parallel dashed lines are reddening
vectors with the crosses placed at intervals corresponding to five
magnitudes of visual extinction.  We have assumed the interstellar
reddening law of \citet{rieke85} ($A_J/A_V$=0.282; $A_H/A_V$=0.175 and
$A_K/A_V$=0.112). The plot is classified into three regions: cool
giants, normally reddened stars and infrared excess sources. 
The numbers correspond to the numbered sources of Fig. \ref{sourcesfig} (a) - (c).}
\label{CCD}
\end{figure}

\begin{figure}
\centering
\includegraphics[totalheight=0.25\textheight,viewport=0 0 700
470,clip]{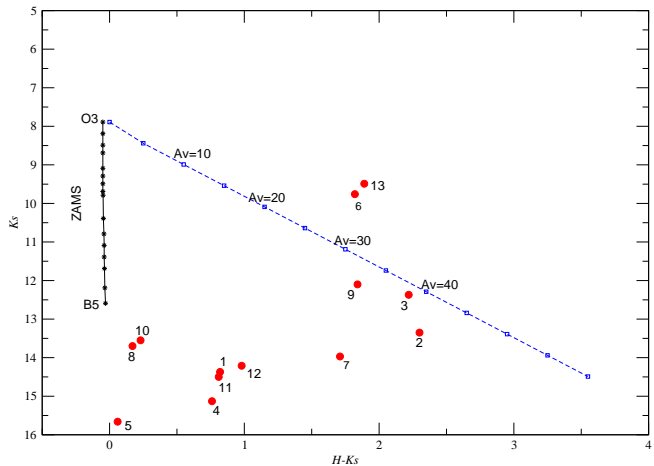}
\caption{Color-magnitude diagram of the 13 2MASS infrared sources towards
IRAS 18544+0112. The solid curve represents the position of the
main sequence at a distance of 3 kpc. The reddening vector for an O3
star, with the squares placed at intervals corresponding to five
magnitudes of visual extinction, is shown with a dashed line. 
The numbers correspond to the numbered sources of Fig. \ref{sourcesfig} (a) - (c).}
\label{CMD}
\end{figure}

Figures \ref{CCD} and \ref{CMD} display the ({\it H-Ks}) versus ({\it
J-H}) color-color (CC) diagram and the ({\it H-Ks}) versus {\it Ks}
color-magnitude (CM) diagram, respectively, of the 13 selected 2MASS
sources.

Figure \ref{CCD} shows that sources one to four, seven and twelve lie in the region of sources with infrared excess. In particular
sources two and three appear as the most reddened ones. Sources six,
nine, eleven and thirteen correspond to normally reddened main sequence stars
and sources eight and ten, the most conspicuos ones in the optical
image (Fig. \ref{sourcesfig}c), are probably no reddened foreground
stars.

According to the CM diagram (Fig. \ref{CMD}), sources two, three, and
nine are the earliest spectral type stars in the region. Sources six
and, thirteen lie in the region of the giant stars, and sources one, four,
five, seven, eight, ten, eleven, and twelve correspond to spectral-type stars
later than B3.  Moreover, as Figure \ref{sourcesfig} (c) shows,
sources one, five, eight, and ten are detected in the DSS2-B optical
band which, given the high obscuration of the region, suggests that
they are probably foreground stars.

Thus, based on both diagrams, we conclude that the most likely
high mass protostars related to \ir\, would be sources two and
three. The fact that source three lies towards the geometrical center of
the IR nebula and that it is placed on a maximun of the 24 $\mu$m
emission while source two is located onto the nebulas's border,
suggests that source three is the main one responsible for the observed
infrared nebula.

\section{Possible scenario}

In this Section we discuss a possible formation scenario for
IRAS~18544+0112.

As mentioned in Sect. 2, the presence of a PDR bordering the HII
region G034.8$-$0.7 together with the observed morphology of its
associated molecular cloud, are clear evidence of the fact that the
HII region  perturbs its enviroment.

\begin{figure*}
\centering
\includegraphics[width=12cm]{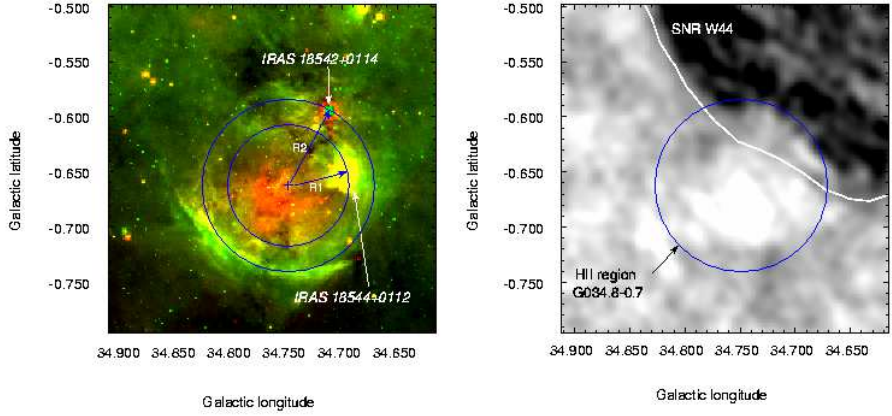}
\caption{{\it Left}: Two-color {\it Spitzer} image of the HII
region G034.8$-$0.7 (8 $\mu$m = green and 24 $\mu$m = red). The
exterior and interior circles correspond to the size of the HII region
at the moment it reached the location of IRAS~18542+0114 and \ir\,,
respectively. {\it Right}: Tomographic map obtained between 74 and 324
MHz (50 \arcsec resolution). Black regions correspond to negative radio
spectral indices, while greys and white indicate regions with a
positive spectrum. Overlapped there is 1.4 HGz radio continuum
emission contour traced at 40 K. The circle is the exterior one from
the left pannel.}
\label{8+24.}
\label{escenario}
\end{figure*}

The IR source \ir\, is seen in projection inside the HII region
G034.8$-$0.7, while IRAS~18542+0114 \citep{paron09} is located upon
its border (see Fig. \ref{escenario}). Based on an infrared and
molecular study, \citet{paron09} found that IRAS~18542+0114 is a MYSO
with molecular outflow activity and whose formation was probably
triggered by the expansion of G034.8$-$0.7 onto the molecular cloud.
In this context, we suggest that the HII region G034.8$-$0.7 has also
triggered the formation of \ir.

The observational differences found between both IRAS sources as
well as their relative location with respect to the center of the HII
region G034.8$-$0.7, suggest that the associated shockfront reached
first the molecular gas where IRAS~18544+0112 was formed, thus explaining
their different evolutionary stages. \citet{koba08} proposed a similar
scenario for the star forming region in Digel's Cloud 2 , where an
expanding shell has perturbed different parts of a molecular cloud
during its evolution, producing different star generations.

To test the proposed scenario we estimate the age of the HII region at
the moment it reached the location of each IRAS source. Using a simple
model described by \citet{dys80} we calculated  the age of the HII
region at a given radius $R$ as

$$t(R)=\frac{4~R_s}{7~c_s}\left[\left(\frac{R}{R_s}\right)^{7/4}-1\right]$$

where $c_s$ is the sound velocity in the ionized gas  ($c_s$=10
\k) and $R_s$ is the radius of the Str\"omgren sphere, given by
$R_s=(3N_{uv}/4\pi n_0^2\alpha_B)^{1/3}$, where $\alpha_B$=2.6 $\times
10^{-13}$ cm$^3$ s$^{-1}$ is the hydrogen recombination coefficient to
all levels above the ground level. $N_{uv}$ is the total number of
ionizing photons per unit of time emitted by the stars, and $n_0$ is
the original ambient density.

The number $N_{\mathrm{uv}}$ of UV ionizing photons needed to keep an
 HII region ionized is given by \citep[see, e.g.][]{cha76}
 $N_{\mathrm{uv}}= 0.76 \, \times 10^{47}\, T_4^{-0.45}\, \nu_{\rm
 GHz}^{0.1}\,S_{\nu}\, D_{\rm kpc}^2 $, where $T_{\mathrm{4}}$ is the
 electron temperature in units of $10^4$~K, $D_{\rm kpc}$ the distance
 in kpc, $\nu_{\rm GHz}$ the frequency in GHz, and $S_{\nu}$ the
 measured total flux density in Jy.  To calculate the radio flux
 density and in order to overcome the problem of superposition of the
 thermal radiation of the HII region and the non-thermal flux of the
 SNR that occurs in the western half of the HII region, we assumed
  the HII region to consist of two identical halves; we therefore
 integrated the flux density over the ``free'' eastern half of the
 ionized sphere and assumed that for the complete region it is just
 twice this value.  Using the 1.4 GHz VLA image we estimated
 $S_{\mathrm{1.4 GHz}}$ $\sim$ 3~Jy, which  agrees with
 previous estimations, $S_{\mathrm{4.85 GHz}}$ = 2.6~Jy \citep{kuc97},
 and $S_{\mathrm{2.7 GHz}}$ = 2.8~Jy \citep{pal03} for a spectral
 index $\alpha$ = $-$0.1 ($S_{\nu} \sim \nu^{\alpha}$) typical for the
 optically thin regime of HII regions.  For an HII region with T=
 10$^{4}$~K placed at a distance of 3~kpc, the total amount of
 ionizing photons needed to keep the source ionized turns out to be
 about $N_{\mathrm{uv}}= 1.7 \times 10^{48} \rm ph\,s^{-1}$.  Based on
 the ionizing fluxes for massive stars given by \citet{sch97}, we
 infer that the ionizing star cannot be later than O9.5\,V.  However,
 this is only a coarse limit. The exciting star is probably earlier
 than O9.5 since if the observed infrared emission originates in dust
 heated by stellar radiation, part of the UV radiation is dissipated
 in this way.

To estimate the dynamical age of the HII region it is important to
determine the center from which the ionized gas is expanding.  Based
on the infrared and radio continuum emission distribution we identify
an almost circular morphology for the HII region G034.8$-$0.7. Figure
\ref{escenario}\,(left) shows that infrared emission at 8 $\mu$m
(green) clearly delineates the southeastern border of the HII region,
while the border where IRAS 18542+0114 is located is diffuse and
fainter. To better delineate this border we inspected a tomographic
map traced between 74 and 324 MHz towards W44 (see
Fig. \ref{escenario}º,(right)). In a radio study of SNR W44,
\citet{cast07} found a curious spectral index inversion on the
southeastern limb of W44 that appears as an indentation in the SNR
boundary. There, the spectrum changes from a negative value
corresponding to the SNR synchrotron radiation (spectral index $\alpha
\sim -0.6$) to $\alpha \geq 0$. In Fig. \ref{escenario}\,(right) darker
regions correspond to a negative spectral index. We propose that the
spectral inversion must be the result of free-free absorption produced
by thermal ionized gas naturally explained by the presence of the HII
region G034.8$-$0.7 located between us and the SNR. In this way, we
can delineate the northwestern border of the HII region where IRAS
18542+0114 is located, thus determining the probable center of the
expanding HII region.

By adopting the radii R$_1$=3pc and R$_2$=4pc for \ir\, and
IRAS~18542+0114, respectively, an ambient density in the region of less
than $\sim$ 10$^{3}$~cm$^{-3}$ and as the exciting agent for the HII
region an O9.5 star, we derived dynamical ages of about  0.6 $\times
10^6$ and 1.2 $\times 10^6$ yr for the HII region at the position of
\ir\, and IRAS~18542+0114, respectively. An age difference of $\sim$  6
$\times 10^5$ yr is compatible with the observed different evolutive
stages of both IR sources.

The remaining question is whether the expansion of the SNR W44 had
some influence in the processes observed in this molecular complex.
Based on the age of the associated pulsar, \citet{wol91} estimated an
age of about $2 \times 10^{4}$ yr for the SNR W44.  Thus, the
possibility that the SNR W44 has triggered the star formation in this
region seems unlikely.

\section{Summary}

We present a molecular and infrared analysis of the IR source
\ir.  The main results can be summarized as follows:

(a) We find \2 J=3--2 self-absorbed profiles, which are typical of
star forming regions. However, we do not detect any evidence of
an outflow activity in \ir\, either in the molecular lines or in the
infrared emission distribution.

(b) The analysis of the \2 J=3--2 line suggests the presence of
expanding molecular gas in the region. 

(c) Based on its morphology, infrared and molecular parameters, and
the non detection of ionized gas, we suggest that \ir\, is an evolved
high mass protostellar object which has not yet reached the ultracompact
HII region stage.

(d) From a near-infrared photometric analysis of the point sources
observed towards \ir\, we propose that 2MASS 18565878+0116233 source
(source three in the text) is the MYSO candidate most likely associated
with \ir.

(e) Based on the observational evidence that the HII region
G034.8$-$0.7 is perturbing the neighboring molecular clouds we suggest
that it has triggered at least two star forming regions, IRAS
18542+0114 and \ir\,, during its expansion.

\begin{acknowledgements}
S.P. is grateful to the staff of ASTE for the support received during
the observations, especially to Juan Cort{\'e}s.  S.P. acknowledges
the support of Viviana Guzm{\'a}n during the observations.  M.O. is a
doctoral fellow of CONICET, Argentina. S.P., S.C., G.C., and G.D. are
members of the {\sl Carrera del Investigador Cient\'\i fico} of
CONICET, Argentina. This work was partially supported by Argentina
grants awarded by CONICET, UBA and ANPCYT.  M.R. is supported by the
Chilean {\sl Center for Astrophysics} FONDAP No. 15010003. M.R. and
S.P. acknowledge support from FONDECYT N\d~1080335. This work has made
use of GLIMPSE and MIPSGAL data obtained with the Spitzer Space
Telescope, which is operated by the Jet Propulsion Laboratory,
California Institute of Technology, under NASA contract 1407. We also
used data products from the Two Micron All Sky Survey, the NASA/IPAC
Infrared Science Archive, which is operated by the Jet Propulsion
Laboratory, California Institute of Technology, under contract with
NASA. We are grateful to the anonymous referee, whose comments and
suggestions led to the improvement of this Paper.

\end{acknowledgements}

\bibliographystyle{aa}  
\bibliography{bibliografia-N}
   
\IfFileExists{\jobname.bbl}{}
{\typeout{}
\typeout{****************************************************}
\typeout{****************************************************}
\typeout{** Please run "bibtex \jobname" to optain}
\typeout{** the bibliography and then re-run LaTeX}
\typeout{** twice to fix the references!}
\typeout{****************************************************}
\typeout{****************************************************}
\typeout{}
}

\end{document}